\documentclass[aps,prb,showpacs]{revtex4}
\usepackage{graphicx, amsmath, amsfonts, amssymb}

\begin{document}

\title{Aharonov-Bohm quantum rings in high-Q microcavities}

\author{A. M. Alexeev}
\affiliation{School of Physics, University of Exeter, Stocker Road, Exeter EX4 4QL, United Kingdom}

\author{I. A. Shelykh}
\affiliation{Division of Physics and Applied Physics, Nanyang Technological University 637371, Singapore}
\affiliation{Science Institute, University of Iceland, Dunhagi 3, IS-107, Reykjavik, Iceland}

\author{M. E. Portnoi}
\email[]{m.e.portnoi@exeter.ac.uk}
\affiliation{School of Physics, University of Exeter, Stocker Road, Exeter EX4 4QL, United Kingdom}
\affiliation{International Institute of Physics, Av. Odilon Gomes de Lima, 1722, Capim Macio, CEP: 59078-400, Natal - RN, Brazil}

\date{February 8, 2013}

\begin{abstract}
A single-mode microcavity with an embedded Aharonov-Bohm quantum ring, which is pierced by a magnetic flux and subjected to a lateral electric field, is studied theoretically. It is shown that external electric and magnetic fields provide additional means of control of the emission spectrum of the system. In particular, when the magnetic flux through the quantum ring is equal to a half-integer number of the magnetic flux quantum, a small change in the lateral electric field allows tuning of the energy levels of the quantum ring into resonance with the microcavity mode providing an efficient way to control the quantum ring-microcavity coupling strength. Emission spectra of the system are calculated for several combinations of the applied magnetic and electric fields.
\end{abstract}


\maketitle

\section{Introduction}

\label{Inroduction}

Progress in nanolithography and epitaxial techniques has resulted in burgeoning developments in the fabrication of micro-scale optical resonators, known as optical microcavities. If a quality factor of a cavity is sufficiently large, the formation of hybrid light-matter excitations occurs. Being observed for the first time two decades ago, \cite{Weisbuch} the strong coupling regime is now routinely achieved in different kinds of microcavities. \cite{Microcavities} From the point of view of fundamental physics this regime is interesting for investigation of various collective phenomena in condensed matter systems such as the high-temperature Bose-Einstein condensation (BEC) \cite{BEC} and superfluidity. \cite{Superfluidity} From the viewpoint of applications it opens a way forward to the realization of optoelectonic devices of a new generation: \cite{OptApplications} room-temperature polariton lasers \cite{PolLasers}, polarization- controlled optical gates, \cite{OptGates} effective sources of THz radiation, \cite{THz} and others.

Several applications of the strong coupling regime were also proposed for quantum information processing. \cite{QInformatics,Johne} In this case one should be able to tune the number of emitted photons in a controllable way. This is hard to achieve in planar microcavities, where the number of elementary excitations is macroscopically large, but is possible in microcavities containing single quantum dots (QDs), where the QD exciton can be coupled to a confined electromagnetic mode provided by a micropillar (etched planar cavity), \cite{QDStrongCouplingPillar} a defect of the photonic crystal, \cite{QDStrongCouplingDefect} or a whispering gallery mode. \cite{Peter2005,Kaliteevski2007} That is why the strong coupled systems based on QDs have attracted particular attention recently. In the strong coupling regime the system possesses a rich multiplet structure, which maps transitions between quantized dressed states of the light-matter coupling Hamiltonian. \cite{QDStrongCouplingPillar, QDStrongCouplingDefect, Peter2005, Hennessy2007, Laussy2008, Laussy2009, Valle2009, Valle2010, Savenko2012}

On the other hand, there is a considerable interest in non-simply-connected nanostructures, such as quantum rings (QRs), which have been obtained in various semiconductor systems. \cite{Lorke2000,Ribeiro2004, Chen2009} This interest is caused by a wide variety of purely quantum mechanical topological effects which are observed in ring-like nanostructures and are absent in quantum dots. The star amongst them is the Aharonov-Bohm effect, \cite{Siday1949, Aharonov1959} in which a phase of a quantum particle is influenced by a vector potential which results in magnetic-flux-dependent oscillations of the particle energy. Recently it was shown that an external lateral electric field, which is known to reduce the QR symmetry and suppress the energy oscillations for the low-energy states, \cite{Barticevic2002, BrunoAlfonso2005} also modifies optical properties of the QR. \cite{Fischer2009,GonzalesSantander2011,1Alexeev2012, 2Alexeev2012} Namely, the application of a weak electric field leads to magneto-oscillations of the degree of polarization of optical transitions between the ground and the first excited states, which are typically in the THz range. When the
magnetic flux through the QR is equal to a half-integer number of flux quanta, these transitions are linearly polarized with the polarization vector normal to the direction of the external electric field, and their frequencies are completely controlled by the magnitude of the applied electric field. \cite{1Alexeev2012,2Alexeev2012} This provides additional means of tuning the QR emission spectrum.

\begin{figure}[h]
\centering
\includegraphics*[width=0.4\linewidth]{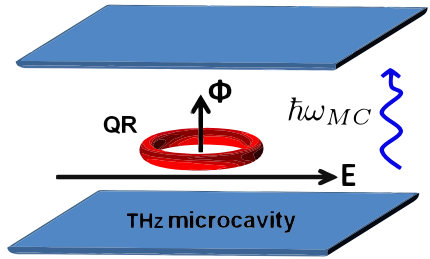}
\caption{An Aharonov-Bohm quantum ring embedded into a single-mode THz microcavity.}
\label{QR_in_MC}
\end{figure}

In the present work we examine a single-mode THz microcavity
\cite{THz_microcavities,Yee2009} with an embedded
Aharonov-Bohm quantum ring, which is pierced by a magnetic flux and
subjected to a lateral electric field. We restrict our analysis to
linearly polarized microcavity radiation only. The geometry of the
system is shown in Fig.~\ref{QR_in_MC}.
The emission properties of such a system under continuous incoherent
pumping are studied theoretically. We calculate the luminescence
spectrum of the system using the master equation techniques for
several combinations of the applied external electric and
magnetic fields. We demonstrate that the resonance, which is best for
exploring quantum features of the system, \cite{Valle2009} can be
achieved by means of tuning the magnitude of the lateral electric field.
An additional degree of control can be achieved by changing the
angle between the polarization plane of the optical pump and the lateral
electric field. As we show, the QR-microcavity coupling strength
depends strongly on the above mentioned angle.

\section{Model}

\subsection{An Aharonov-Bohm quantum ring in an external electric field}

In this section we revise the energy spectrum and optical properties
of a single-electron Aharonov-Bohm QR pierced by a magnetic flux
$\Phi$ and subjected to a lateral electric field $\mathbf{E}$, which
were studied in Refs. \onlinecite{2Alexeev2012} and \onlinecite{1Alexeev2012}.

In the absence of the external electric field the eigenfunctions of an infinitely narrow Aharonov-Bohm QR of a radius $R$ are given by
\begin{equation}
\label{QRwfunction}
\psi_{m} \left ( \varphi \right ) =e^{i m \varphi} / \sqrt{2 \pi} \mbox{,}
\end{equation}
where $\varphi$ is the polar angle coordinate and $m= 0, \pm 1, \pm 2 ...$ is the angular momentum quantum number. The corresponding eigenvalues are defined by
$$
\varepsilon_{m}=\varepsilon_{QR} \left ( m + \phi \right )^{2} \mbox{,}
$$
where $\varepsilon_{QR}=\hbar^{2}/2 M_{e} R^2$ is
the energy scale of the interlevel separation in the QR, $M_{e}$ is the
electron effective mass and $\phi=\Phi / \Phi_{0} $ is the number
of flux quanta piercing the QR ($\Phi_{0}= h / e $). For
experimentally attainable QRs, $\varepsilon_{QR}$ corresponds to the THz frequency range.
\cite{1Alexeev2012}

When the lateral electric field is applied, the modified electron eigenfunctions can be expressed as a linear combination of the unperturbed wave functions (\ref{QRwfunction}):
\begin{equation}
\label{QRmwfunction}
\Psi \left ( \varphi \right )= \sum \limits_{m} C_{m} e^{i m \varphi} \mbox{.}
\end{equation}
Substituting the wave function (\ref{QRmwfunction}) into the Schr\"{o}dinger equation with the Hamiltonian containing a term which describes the presence of the lateral electric field, multiplying the resulting expression by $e^{-i m \varphi}$, and integrating with respect to the angle $\varphi$ results in an infinite system of linear equations for the coefficients $C^{n}_{m}$
\begin{equation}
\left [ \left ( m+ \phi \right ) ^{2} - \Lambda \right ] C_{m} + \beta \left ( C_{m+1} + C_{m-1} \right ) = 0\,\mbox{,}
\label{Linearsystem}
\end{equation}
where $\beta=eER/2\varepsilon_{QR}$ is the normalized strength of the lateral electric field and $\Lambda$ is an energy eigenvalue normalized by $\varepsilon_{QR}$. It can be seen from the system of equations (\ref{Linearsystem}) that all the QR quantities are periodic in the magnetic flux $\Phi$ with the period equal to $\Phi_{0}$. There is also an apparent symmetry with respect to the change of the sign of $\Phi$. Therefore, in what follows we will consider only the case of $0 \le \Phi \le \Phi_{0}/2$.

\begin{figure}[h]
\centering
\includegraphics*[width=0.4\linewidth]{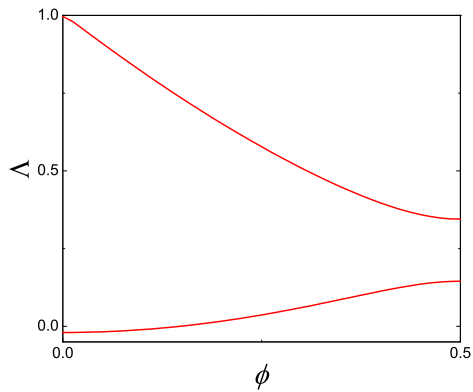}
\caption{The normalized energy spectrum for the electron ground and the first excited states in the QR as a function of dimensionless parameter $\phi$ for $\beta=0.1$.}
\label{QREnergySpectrum}
\end{figure}

It was shown in Ref.~\onlinecite{1Alexeev2012} that in the limit of a weak in-plane electric field, $eER \ll \varepsilon_{QR}$, all essential features of the first three states of the QR are fully captured by the following $3 \times 3$ system of linear equations:
\begin{equation}
\label{33system}
\begin{pmatrix} \left ( \phi+1 \right )^{2} & \beta & 0 \\ \beta & \phi^{2} & \beta \\ 0 & \beta & \left ( \phi-1 \right )^{2} \end{pmatrix} \begin{pmatrix} C_{+1} \\ C_{0} \\ C_{-1} \end{pmatrix} = \Lambda \begin{pmatrix} C_{+1} \\ C_{0} \\ C_{-1} \end{pmatrix} \mbox{.}
\end{equation}
In what follows we will be interested in the transitions between the ground and the first excited states in the ring only. However, in order to obtain accurate ground and first excited states eigenenergies and eigenfunctions all three listed states should be considered (for details see Ref.~\onlinecite{1Alexeev2012}). The system of linear equations (\ref{33system}) can be reduced to a cubic equation for $\Lambda$, which yields the following eigenvalues:
\begin{equation}
\label{lambda1}
\Lambda_{1} = - 2/3 \sqrt{1+ 12 \phi^{2} +6 \beta^2} \cos \left( \alpha /3 \right) + \phi^{2}+2/3 \mbox{,}
\end{equation}
\begin{equation}
\label{lambda2}
\Lambda_{2} = - 2/3 \sqrt{1+ 12 \phi^{2} +6 \beta^2} \cos \left( \alpha /3 - 2 \pi /3 \right ) + \phi^{2}+2/3 \mbox{,}
\end{equation}
\begin{equation}
\label{lambda3}
\Lambda_{3} = - 2/3 \sqrt{1+ 12 \phi^{2} +6 \beta^2} \cos \left( \alpha /3 + 2 \pi /3 \right ) + \phi^{2}+2/3 \mbox{,}
\end{equation}
where
$$
\cos{\alpha}=\frac { 1 - 36 \phi^2 + 9 \beta^{2}}  { \left (1 + 12 \phi^{2} +6 \beta^2 \right )^{3/2}} \mbox{.}
$$
The set of corresponding eigenvectors (non-normalized) is given by substituting appropriate values of $\Lambda$ into
\begin{equation}
\label{mwf_coeffitients}
\begin{pmatrix} C_{+1} \\ C_{0} \\ C_{-1} \end{pmatrix} = \begin{pmatrix} \left [ \Lambda  - \left( \phi -1 \right)^{2} \right] \left ( \Lambda-\phi^2 \right ) - \beta^{2} \\ \left [ \Lambda - \left( \phi-1 \right)^{2} \right]  \beta \\ \beta^{2} \end{pmatrix} \mbox{.}
\end{equation}
The energy spectrum for the electron ground and the first excited
states defined by Eq.~(\ref{lambda1}) and Eq.~(\ref{lambda2}) for
$\beta=0.1$ and $0 \le \phi \le 1/2$ is plotted in
Fig.~\ref{QREnergySpectrum}.
Notably, the $3\times3$ system of equations (\ref{33system}) provides a very good accuracy for the ground and the first excited states when $\beta \lesssim 1$ ($eER \lesssim \varepsilon_{QR}$). A numerical check shows that the further increase in the system of linear equations, Eq.~(\ref{Linearsystem}), does not provide any noticeable change in the results.
A similar analysis is applicable to a nanohelix with an electric field applied normal to its axis. For a helix, the role of magnetic flux in the absence of a magnetic field is played by the electron momentum along the helical line. \cite{helix0,helix1, helix2, helix3}

Another quantity, which is needed for our further calculations, is the product of the light polarization vector $\mathbf{p}$ and the matrix element $\mathbf{d}= \left \langle e \left| \hat{\mathbf{d}} \right| g \right \rangle = \left \langle g \left| \hat{\mathbf{d}} \right| e \right \rangle $ of the dipole moment calculated between the ground state $\left| g \right \rangle$ and the first excited state $\left| e \right \rangle$. For linearly polarized light this product is given by the following integral:
\begin{equation}
\label{tmatrixelement_gf} \mathbf{d} \cdot \mathbf{p}=eR
\int_0^{2\pi} \Psi_{e} \Psi_{g} \cos \left ( \theta - \varphi
\right) d \varphi \mbox{,}
\end{equation}
where $\Psi_{g}$, $\Psi_{e}$ are the ground and the first excited state wave functions defined by Eq.~(\ref{QRmwfunction}) and $\theta$ is the angle between $\mathbf{p}$ and $\mathbf{E}$.

Substituting eigenfunctions $\Psi_{g}$, $\Psi_{e}$ given by Eq.~(\ref{QRmwfunction}) into Eq.~(\ref{tmatrixelement_gf})
and performing the integration with respect to the angle $\varphi$ we obtain
\begin{equation}
\label{tmatrixelement_ef}
\mathbf{d} \cdot \mathbf{p} = \left(d_{-}^{2}+d_{+}^2-2d_{-} d_{+} \cos 2 \theta \right )^{1/2} \mbox{,}
\end{equation}
where
\begin{equation}
\label{d_m}
d_{-}= \frac {e R} {2} \left |C_{0}^{e} C_{-1}^{g}+C_{+1}^{e} C_{0}^{g} \right| \mbox{,}
\end{equation}
and
\begin{equation}
\label{d_p}
d_{+}= \frac {e R} {2} \left |C_{-1}^{e} C_{0}^{g}+C_{0}^{e} C_{+1}^{g} \right| \mbox{.}
\end{equation}
Later in this paper we use Eqs.~(\ref{tmatrixelement_ef})--(\ref{d_p}) with coefficients $C^{e}$, $C^{g}$ obtained from Eq.~(\ref{mwf_coeffitients}) to calculate the QR-microcavity coupling strength. A detailed analysis \cite{1Alexeev2012} of Eq.~(\ref{mwf_coeffitients}),(\ref{d_m})--(\ref{d_p}) shows that a noticeable $\theta$-dependence in Eq.~(\ref{tmatrixelement_ef}) occurs only when $\phi=0$ or $\phi=1/2$, as $d_{-}$ vanishes otherwise.

\subsection{The Jaynes-Cummings Hamiltonian and the Master Equation}
We represent the QR as a two-level system with the energy gap between
the ground state $\left |g \right \rangle$ and the excited state
$\left |e \right \rangle$ denoted by
$\Delta$. From Eqs.~(\ref{lambda1})--(\ref{lambda2}), it is clear that
$\Delta$ depends on both the external electric field $\mathbf{E}$,
applied in the QR plane, and the magnetic flux $\Phi$, piercing the QR. In
particular, when $\phi=0$ ($\phi=1/2$), one obtains $\Delta /
\varepsilon_{QR}=1 + 2\beta^{2}$ ($\Delta / \varepsilon_{QR}= 2
\beta$). The full Hamiltonian describing the system of a QR
coupled to a single-mode THz microcavity is the Jaynes-Cummings
Hamiltonian \cite{JaynesCummings1963}
\begin{equation}
\label{JCHamiltonian}
H= \Delta \sigma^{\dagger} \sigma + \hbar \omega_{MC} a^{\dagger} a + {\cal G} \left (\sigma^{\dagger} a + \sigma a^{\dagger} \right) \mbox{,}
\end{equation}
where $\omega_{MC}$ is the microcavity eigenfrequency, $\cal G$ is
the QR-microcavity coupling constant, $a^{\dagger}$ is the microcavity photon
creation operator, $a$ is the microcavity photon
annihilation operator, $\sigma^{\dagger}=(\sigma_{x} + i \sigma_{y})/2$ is the
QR electron creation operator, $\sigma=(\sigma_{x} - i \sigma_{y})/2$ is the
QR electron annihilation operator, and $\sigma_{x}$, $\sigma_{y}$ are the
Pauli matrices acting in the space of $\left | g \right\rangle$ and
$\left | e \right\rangle$ states. The frequency of the microcavity
mode and the frequency of the transition between the QR states are assumed to be close
enough to allow the use of the rotating wave approximation.
\cite{Scully, CohenTannoudji,Valle} If the cavity mode is linearly
polarized, $\cal G$ is given by
\begin{equation}
\label{cc_g}
{\cal G} = - \left ( \mathbf{d} \cdot \mathbf{p} \right ) \sqrt{\hbar \omega_{MC} / 2 \epsilon_{0} V} \mbox{,}
\end{equation}
where $\mathbf{d} \cdot \mathbf{p}$ is given by
Eq.~(\ref{tmatrixelement_ef}), $\epsilon_{0}$ is the vacuum
dielectric permittivity, $V$ is the quantization volume, which can
be estimated as $V \approx \left ( \lambda_{MC} /2 \right)^{3}$, and
$\lambda_{MC}=2 \pi c/ \omega_{MC}$ is the microcavity characteristic
wavelength. When the magnetic flux piercing the QR is equal to an
integer number of half-flux quanta, $\cal G$ strongly depends on the
angle $\theta$ between the projection of the radiation polarization
vector onto the QR plane and the applied lateral electric field.

The eigenvalues of the Hamiltonian (\ref{JCHamiltonian}) are the
same as in the case of a single-mode microcavity with embedded QD,
whose excitations obey fermionic statistics \cite{Valle2009,Scully,Valle}
\begin{equation}
\label{JCeigenvalues}
E_{N}^{\pm}=\hbar \omega_{MC} \left (N - 1/2 \right) + \Delta/2 \pm \sqrt{\left ( \hbar \omega_{MC} - \Delta \right )^{2}/4 + N {\cal G}^{2}} \mbox{,}
\end{equation}
where $N$ is the total number of electron-photon excitations
in the system, i.e. the number of photons inside the microcavity
if the electron is in the ground state. The corresponding
eigenfunctions can be expressed as a linear combination of the
combined electron-phonon states $\left| g, N \right \rangle= \left|g
\right \rangle \times \left| N \right \rangle$ and $\left|e, N-1 \right \rangle= \left|e
\right \rangle \times \left| N-1 \right \rangle$, which define both the QR state
and the microcavity photon occupation number. Explicitly, the
eigenfunctions are as follows
\begin{equation}
\label{JCeigenfunctions}
{\cal X}^{\pm}_{N}= K_{g,N}^{\pm} \left| g, N \right \rangle + K_{e,N}^{\pm} \left| e, N-1 \right \rangle \mbox{,}
\end{equation}
where
\begin{equation}
\label{K_g}
K_{g,N}^{\pm}= \frac {\sqrt{N} {\cal G}} {\sqrt{\left( E_{N}^{\pm} - N \hbar \omega_{MC} \right)^{2} + N {\cal G}^{2}}} \mbox{,}
\end{equation}
and
\begin{equation}
\label{K_e}
K_{e,N}^{\pm}= \frac {E_{N}^{\pm} - N \hbar \omega_{MC}} {\sqrt{\left( E_{N}^{\pm} - N \hbar \omega_{MC} \right)^{2} + N \hbar {\cal G}^{2}}} \mbox{.}
\end{equation}
The main advantage of using a QR instead of a QD is the opportunity
to control both the energy gap $\Delta$ between the first two states
of the QR and the QR-microcavity coupling constant ${\cal G}$ by changing the
external electric and magnetic fields. These fields can be used to
achieve the resonant condition $\Delta = \hbar \omega_{MC}$ and provide easy means of performing a transition from the
strong to the weak coupling regime within the same system. \cite{Valle2009}

The eigenvalues $E_{N}^{\pm}$ defined by Eq.~(\ref{JCeigenvalues})
form the so-called ``Jaynes-Cummings ladder'' and the emission
spectrum of the system, which is observed outside of the microcavity,
is defined by optical transitions between the states with total number
of electron-photon excitations $N$ different by unity. Inside a
non-ideal microcavity, a photon has a limited lifetime and when the
photon leaks out, one can measure its frequency. This provides a
direct access to the quantized coupled electron-photon states of the
system.

In order to describe any realistic experiment measuring the
QR-microcavity emission spectrum one should introduce pump and decay
in the system. We model the system dynamics under incoherent cavity
pumping and account for dissipation processes using the master
equation approach for the full density matrix of the system $\rho$
(see, e.g., Refs. \onlinecite{Scully, CohenTannoudji,Valle}). The master equation reads
\begin{equation}
\label{master_equation}
\frac {\partial \rho} {\partial t}= \frac {i} {\hbar} [\rho, H] + {\cal L}_{P}^{MC} \rho + {\cal L}_{\gamma}^{MC}  \rho + {\cal L}_{\gamma}^{QR} \rho \mbox{,}
\end{equation}
where ${\cal L}_{P}^{MC}$, ${\cal L}_{\gamma}^{MC}$ are the Lindblad terms, which account for the microcavity pump and decay, and the Lindblad term ${\cal L}_{\gamma}^{QR}$ describes non-radiative transitions of the QR electron from the excited state $\left| e \right \rangle$ to the ground state $\left| g \right \rangle$. In the explicit form these three terms are given by
\begin{equation}
\nonumber
{\cal L}_{P}^{MC} \rho=\frac {P_{MC}} {2} ( 2 a^{\dagger} \rho a - a a^{\dagger} \rho - \rho a a^{\dagger} + 2 a \rho a^{\dagger} - a^{\dagger} a \rho - \rho a^{\dagger} a) \mbox{,}
\end{equation}
\begin{equation}
\nonumber
{\cal L}_{\gamma}^{MC} \rho=\frac {\gamma_{MC}} {2} ( 2 a \rho a^{\dagger} - a^{\dagger} a \rho - \rho a^{\dagger} a) \mbox{,}
\end{equation}
\begin{equation}
\nonumber
{\cal L}_{\gamma}^{QR} \rho=\frac {\gamma_{QR}} {2} ( 2 \sigma \rho \sigma^{\dagger} - \sigma^{\dagger} \sigma \rho - \rho \sigma^{\dagger} \sigma) \mbox{,}
\end{equation}
where $P_{MC}$ is the intensity of the incoherent microcavity
pumping and $\gamma_{MC}$, $\gamma_{QR}$ are the lifetimes of the
photonic and the QR excitations respectively. Due to the balance between the
pump and the decay, after some time a steady state is established. We denote the
corresponding density matrix as $\rho^{SS}$. The steady
state density matrix can be found by solving numerically
Eq.~(\ref{master_equation}) with all the matrices truncated. When
performing the truncation, all the states which can be excited as
a result of the pumping should be accounted for.

\subsection{Emission spectrum of the system under incoherent pumping}


In the presence of the pump and the decay and after establishing an equilibrium, the system is in a mixed state, which is characterized by the full density matrix $\rho^{SS}$. If $\rho^{SS}$ is written in the basis of eigenfunctions (\ref{JCeigenfunctions}), the density matrix diagonal element $\rho^{SS}_{II}$ gives the probability of the system to be in the $I$th state. At low pumping, $P_{MC} \ll {\cal G}$, and in the case of a high-Q system, $\gamma_{MC},\gamma_{QR} \ll {\cal G}$, which is the best regime to elucidate quantum coupling effects, \cite{Valle2009} the emission spectrum can be calculated using the so-called manifold method, \cite{Valle} which has been proved to provide qualitatively accurate results avoiding heavy numerical calculations (see, e.g., Refs. \onlinecite{Valle, Valle2009, Savenko2012}, and \onlinecite{manifold_proved}). In this approximation the QR and microcavity emission spectra are given by
\begin{equation}
\label{S_QR}
S_{QR} \left( \omega \right) \approx \frac {1} {\pi} \sum \limits_{I,F} \frac{ \left| M^{QR}_{IF} \right|^{2} \rho_{II}^{SS} \Gamma_{IF}} {\left (\hbar \Omega_{IF} - \hbar \omega \right)^2 +\Gamma_{IF}^{2}} \mbox{,}
\end{equation}
\begin{equation}
\label{S_MC}
S_{MC} \left( \omega \right) \approx \frac {1} {\pi} \sum \limits_{I,F} \frac{\left| M^{MC}_{IF} \right|^{2} \rho_{II}^{SS} \Gamma_{IF}} {\left (\hbar \Omega_{IF} - \hbar \omega \right)^2 +\Gamma_{IF}^{2}} \mbox{,}
\end{equation}
where $\bigl| M^{QR}_{IF} \bigr|^2 = \left| \left \langle {\cal X}_{F}, \right| \sigma \left| {\cal X}_{I} \right \rangle \right|^{2}$, $\bigl| M^{MC}_{IF} \bigr|^2 =  \left| \left \langle {\cal X}_{F} \left| a \right| {\cal X}_{I} \right \rangle \right|^{2}$, $\hbar \Omega_{IF} = E^{I}-E^{F}$, ${\cal X}_{i}$ and $ {\cal X}_{f}$ are the QR-microcavity initial and final states eigenfunctions defined by Eq.~(\ref{JCeigenfunctions}), $E^{i}$ and $E^{f}$ are the QR-microcavity initial and final states eigenenergies defined by Eq.~(\ref{JCeigenvalues}), and $\Gamma_{IF}$ is given by
\begin{multline}
\nonumber
\Gamma_{IF}= \frac {\gamma_{QR}} {2} \sum \limits_{J} \left( \bigl| M^{QR}_{JI}  \bigr|^{2} + \bigl| M^{QR}_{JF} \bigr|^{2} \right) + \frac {\gamma_{MC}} {2} \sum \limits_{J} \left( \bigl| M^{MC}_{JI}  \bigr|^{2} + \bigl| M^{MC}_{JF} \bigr|^{2} \right) \\ + \frac {P_{MC}} {2} \sum \limits_{J} \left( \bigl| M^{MC}_{JI}  \bigr|^{2} + \bigl| M^{MC}_{JF} \bigr|^{2} + \bigl| M^{MC}_{IJ}  \bigr|^{2} + \bigl| M^{MC}_{FJ} \bigr|^{2} \right) \mbox{.}
\end{multline}

In Eqs.~(\ref{S_MC})-(\ref{S_MC}) $S_{MC}$ and $S_{QR}$ correspond to photons of two different origins, which can be detected outside of the microcavity by an external observer: the direct emission of the QR and the leaking microcavity photons. In the first case a photon outside of the microcavity is created as a result of the QR electron transition from the excited state $\left| e \right \rangle$ to the ground state $\left| g \right \rangle$ and in the second case the photon is created due to annihilation of a microcavity photon. Substituting ${\cal X}^{\pm}_{N}$ from Eq.~(\ref{JCeigenfunctions}) into the expressions for $\bigl| M_{IF} \bigr|^2$ yields
\begin{equation}
\nonumber
\label{M_QR}
\bigl| M^{QR}_{IF} \bigr|^2 =  \left| K_{g,N_{F}}^{\pm} K_{e,N_{I}}^{\pm} \right|^{2} \delta_{N_{F}, N_{I}-1}  \mbox{,}
\end{equation}
\begin{equation}
\nonumber
\label{M_MC}
\bigl| M^{MC}_{IF} \bigr|^2 =  \left| \sqrt{N_{I}} K_{g,N_{F}}^{\pm} K_{g,N_{I}}^{\pm} + \sqrt{N_{F}} K_{e,N_{F}}^{\pm} K_{e,N_{I}}^{\pm} \right|^{2} \delta_{N_{F}, N_{I}-1} \mbox{.}
\end{equation}
It should be noted that only the transitions between the coupled electron-photon states with the total number of excitations differing by unity are allowed. In the resonant case $\Delta=\omega_{MC}$, for transitions from the $N$th state to the $\left(N-1\right)$th state
\begin{equation}
\label{M_QR_pmmp}
\left| M^{QR}_{\pm \to \mp} \right|^{2}= 1/4 \mbox{,}
\end{equation}
\begin{equation}
\label{M_QR_pmpm}
\left| M^{QR}_{\pm \to \pm} \right|^{2}= 1/4 \mbox{,}
\end{equation}
and
\begin{equation}
\label{M_MC_pmmp}
\left| M^{MC}_{\pm \to \mp} \right|^{2}= \left| \sqrt{N} - \sqrt{N-1} \right|^{2} /4 \mbox{,}
\end{equation}
\begin{equation}
\label{M_MC_pmpm}
\left| M^{MC}_{\pm \to \pm} \right|^{2}= \left| \sqrt{N} + \sqrt{N-1} \right|^{2} /4 \mbox{,}
\end{equation}
with corresponding eigenfrequencies given by
\begin{equation}
\label{ef_outer}
\Omega_{\pm \to \mp}=\omega_{MC} \pm {\cal G} \left(\sqrt{N}+\sqrt{N-1} \right) / \hbar \mbox{,}
\end{equation}
\begin{equation}
\label{ef_inner}
\Omega_{\pm \to \pm}=\omega_{MC} \pm {\cal G} \left(\sqrt{N}-\sqrt{N-1} \right) / \hbar \mbox{.}
\end{equation}
One can see that the observed emission spectrum consists of two symmetric inner peaks at frequencies (\ref{ef_inner}) and two symmetric outer peaks at frequencies (\ref{ef_outer}). Together, these peaks form the so-called ``Jaynes-Cummings fork''. From Eqs.~(\ref{M_QR_pmmp})--(\ref{M_MC_pmpm}) it follows that when the total number of electron-photon excitations in the initial state $N=1$, both $S_{QR}$ and $S_{MC}$ have a shape of the Rabi doublet, and in the case of large excitation numbers, $N\gg1$, $S_{QR}$ is in the form of the Mollow triplet while $S_{MC}$ collapses into a single lasing peak.

\section{Results and Discussion}

In this section we use the formalism which was developed in the previous sections to calculate emission spectra of the QR-microcavity system in the presence of a magnetic flux $\Phi$ piercing the QR and a lateral electric field $\mathbf{E}$. The QR-microcavity system has apparent advantages for exploring the quantum nature of light-matter coupling in nanostructured systems compared to the well-studied QD-based setup. Namely, the parameters of the system can be much easier tuned by external fields. Between all possible combinations of the applied magnetic and electric fields there are two cases of a considerable interest: (a) $\Phi=0$, $\mathbf{p} \perp \mathbf{E}$ and (b) $\Phi=\Phi_{0}/2$, $\mathbf{p} \perp \mathbf{E}$. In both cases, the energy gap between the QR states is tunable by the strength of the lateral electric field. From Eqs.~(\ref{lambda1})--(\ref{lambda2}) we get $\Delta / \varepsilon_{QR} = 1 - 2\beta^{2}$ ($\Delta / \varepsilon_{QR} = 2 \beta$) for $\Phi=0$ ($\Phi=\Phi_{0}/2$). Thus, the energy gap $\Delta$ can be easily adjusted to coincide with the energy of the microcavity mode $\hbar \omega_{MC}$. From Eqs.~(\ref{tmatrixelement_ef})--(\ref{d_p}) and Eq.~(\ref{cc_g}) one can see that when $\Phi=0$ or $\Phi=\Phi_{0}/2$ the QR-microcavity coupling constant ${\cal G}$ strongly depends on the angle $\theta$ between the direction of the external electric field and the projection of the microcavity mode polarization vector onto the QR plane. If $\mathbf{p} \perp \mathbf{E}$, the coupling constant ${\cal G}$ reaches its maximum possible value, and if $\mathbf{p} \parallel \mathbf{E}$, the microcavity mode and the QR are completely uncoupled. By changing the direction of the lateral electric field one acquires additional means of control of the emission spectrum of the system.

The quantum structure of the Jaynes-Cummings states discussed in the previous section is known to be observed only in the low dissipation regime. \cite{Valle2009} Therefore, it is natural to consider a QR embedded into a high-Q THz microcavity under a weak incoherent pumping. Similar to Ref.~[\onlinecite{Valle2009}], we choose a microcavity with the decay rate $\gamma_{MC} / {\cal G}=0.1$ and a QR with the decay rate $\gamma_{QR}/ {\cal G}=0.01$. The QR decay rate is chosen to be much smaller than the microcavity decay rate, as is the case in most experimental systems. \cite{QDStrongCouplingPillar, Hennessy2007} In all the calculations we chose either $P_{MC}/ {\cal G}=0.005$ or $P_{MC}/ {\cal G}=0.095$. These conditions satisfy the applicability criteria of the manifold method for modelling the emission spectrum of the systems.

In order to estimate experimental conditions for the observation of the predicted emission spectra we use the following values of the other system parameters: a typical radius of experimentally attainable \cite{Lorke2000,Ribeiro2004,Chen2009} QRs, $R = 20 \mathrm {nm}$ and the electron effective mass $M=0.05 m_{e}$. This gives the energy scale of the QR interlevel separation $\varepsilon_{QR} \simeq 2 \mathrm{meV}$ and the magnitude of the magnetic field, which produces a magnetic flux through the QR equal to a half of the flux quantum, $B \simeq 2 \mathrm{T}$. Unless specified otherwise, all the calculations are made in the presence of a weak lateral electric field $\mathbf{E} \perp \mathbf{p}$ with the magnitude $E=0.1 \varepsilon_{QR} / e R= 2 \cdot 10^{4} \mathrm{V/m}$. The QR-microcavity coupling constant can be now estimated using Eq.~(\ref{cc_g}). We obtain $ {\cal G} =8.3 \cdot 10^{-4} \mathrm{meV}$ ($ {\cal G} = 1.2 \cdot 10^{-3} \mathrm{meV}$) for $\Phi=0$ ($\Phi=\Phi_{0}/2$) which results in the microcavity Q-factor requirement $Q=\hbar \omega_{MC}/\gamma_{MC}\approx 16000$ ($Q \approx 5000$). THz microcavities with the Q-factor of this order of magnitude have already been achieved. \cite{Yee2009}

\begin{figure}[h]
\includegraphics*[width=0.3\linewidth]{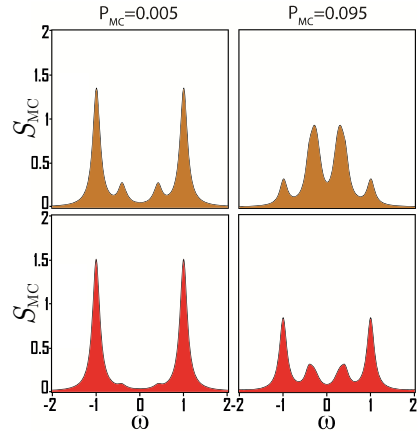}
\caption{(Colour online) Emission spectrum of the QR-microcavity system in the presence of a lateral electric field $E=2.00 \times 10^{4} \mathrm{V/m}$ for $P_{MC}/{\cal G}=0.005$ and $P_{MC}/{\cal G}=0.095$. The microcavity mode is in resonance with the QR transition. The upper row (brown) corresponds to the microcavity emission and the lower row (red) corresponds to the direct QR emission. The magnetic flux piercing the QR is either $\Phi=0$ or $\Phi=\Phi_{0}/2$. The emission frequencies are normalised by the QR-microcavity coupling constant ${\cal G}/ \hbar$ and centred around $\omega_{MC}$.}
\label{ES_pump_change}
\end{figure}

We start with calculations of the emission spectrum of the system for $P_{MC}/ {\cal G}=0.005$ and $P_{MC}/ {\cal G}=0.095$ in the resonant case, $\hbar \omega_{MC}=\Delta$. The magnetic flux piercing the QR is either $\Phi=0$ or $\Phi=\Phi_{0}/2$. The results of our calculations are shown in Fig.~\ref{ES_pump_change}. Both the direct QR emission spectrum, $S_{QR}$, and the microcavity emission spectrum $S_{MC}$ are plotted. When $P_{MC}/ {\cal G}=0.005$, there are two dominant peaks (the linear Rabi doublet) in $S_{QR}$ and $S_{MC}$ at the frequencies $\omega= \pm {\cal G} / \hbar$, which correspond to the transitions between the two $N=1$ states and the ground $N=0$ state. With increasing pumping, $P_{MC}/ {\cal G}=0.095$, the higher, $N>1$, states are excited. The intensity of the Rabi doublet is decreased while the quadruplet peaks corresponding to the transitions between the $N=2$ and $N=1$ states emerge. Only the inner quadruplet peaks in $S_{QR}$ and $S_{MC}$ can be seen in the selected energy range. It should be mentioned that the outer peaks in the microcavity emission spectrum, $S_{MC}$, become suppressed with increasing $N$, as can be seen from the expression for the corresponding matrix elements, Eq.~(\ref{M_MC_pmmp}).

\begin{figure}[h]
\centering
\includegraphics*[width=0.3\linewidth]{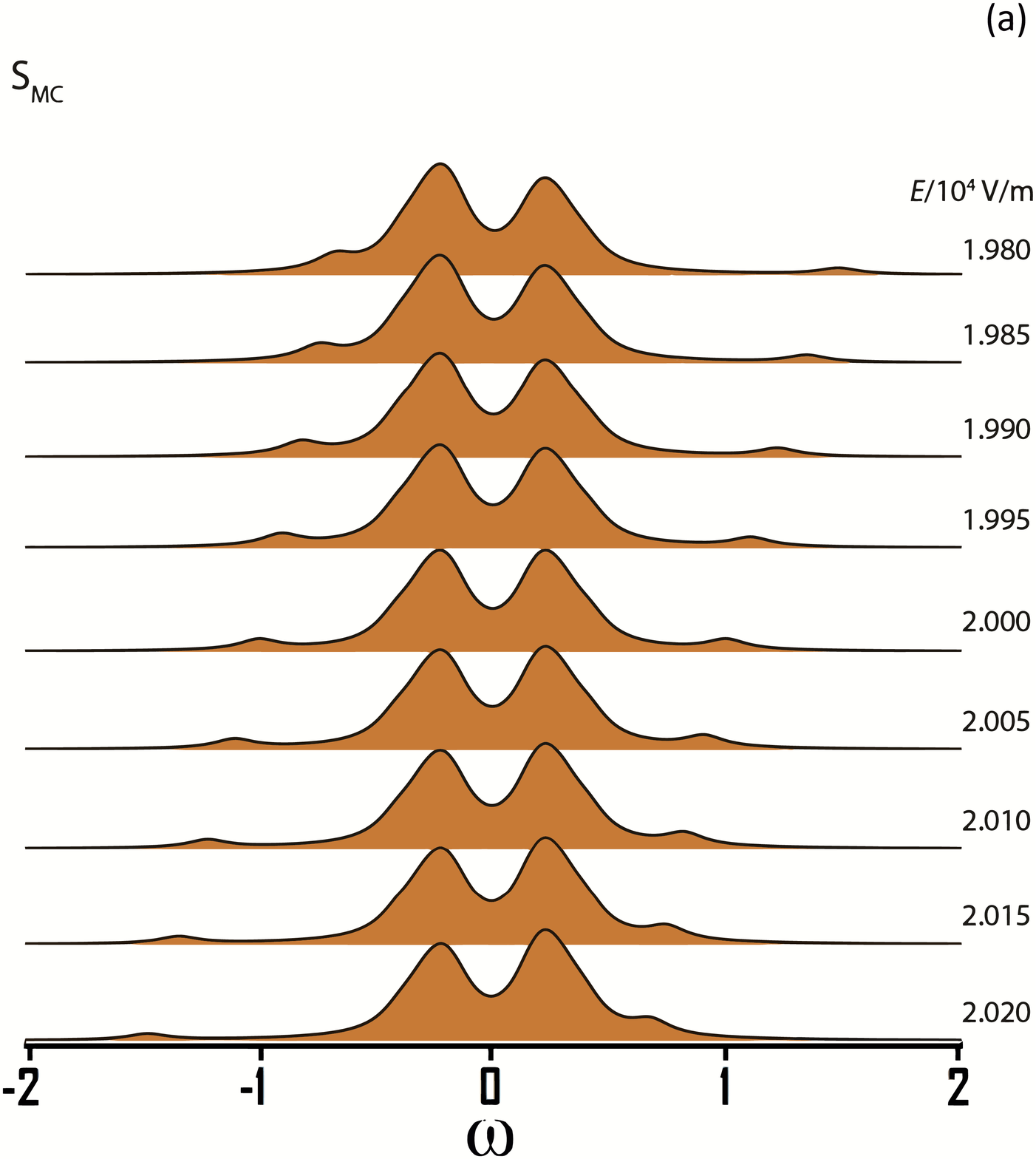}
\includegraphics*[width=0.3\linewidth]{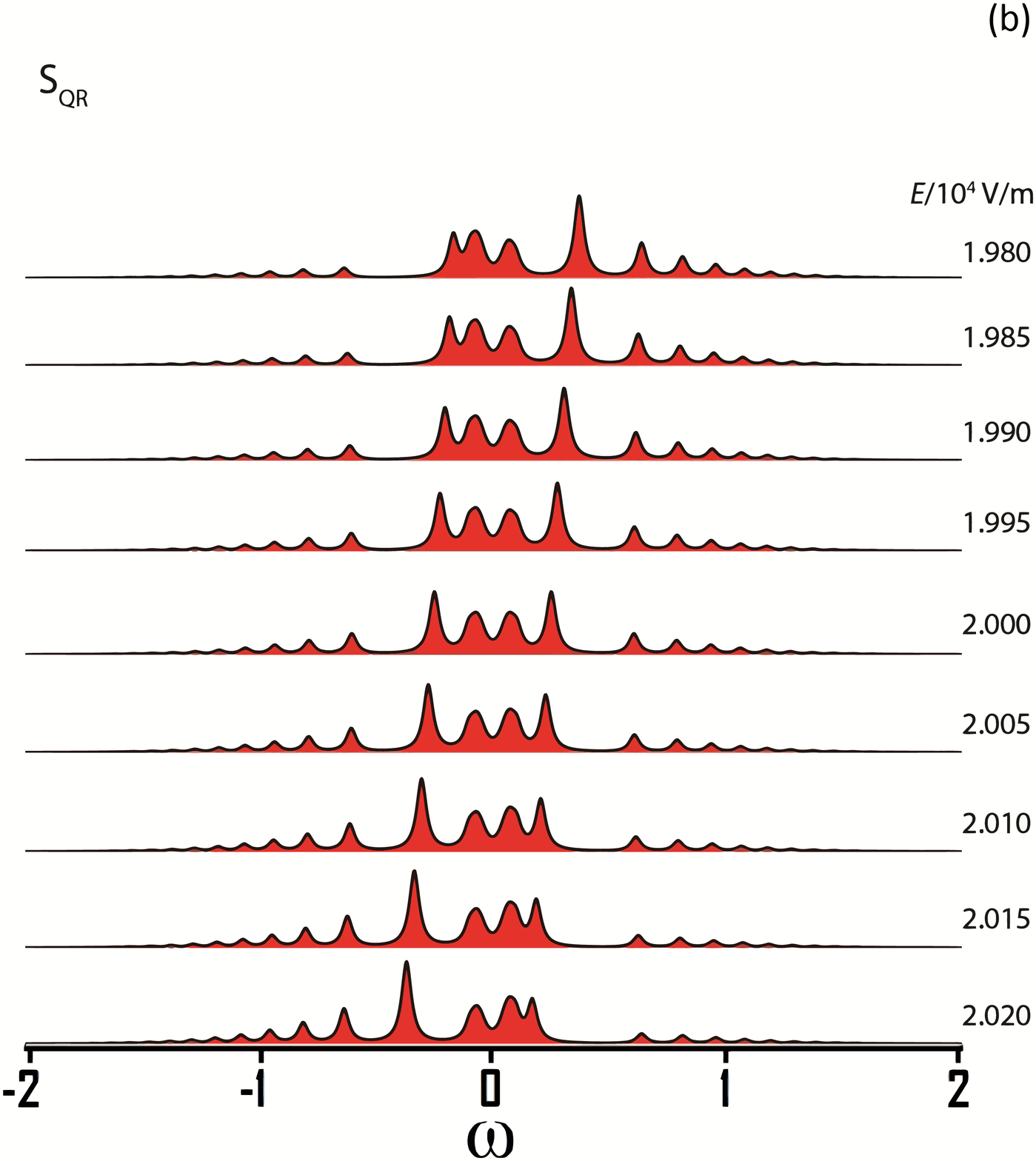}
\caption{(Colour online) Anticrossing in the emission spectrum of the QR-microcavity system at various magnitudes of the external lateral electric field $E$ from $1.98 \times 10^{4} \mathrm{V/m}$ to $2.02 \times 10^{4} \mathrm{V/m}$ with the increment $50 \mathrm{V/m}$: (a) microcavity emission spectrum (brown), (b) direct QR emission spectrum (red). The magnetic flux piercing the QR $\Phi=0$. The resonance case $\Delta=\hbar \omega_{MC}$ corresponds to $E=2.00 \times 10^{4} \mathrm{V/m}$. The microcavity pumping rate $P_{MC}/{\cal G}=0.095$. The emission frequencies are normalised by the QR-microcavity coupling constant ${\cal G}/ \hbar$ and centred around $\omega_{MC}$.}
\label{ES_E_change_F_00}
\end{figure}
\begin{figure}[h]
\centering
\includegraphics*[width=0.3\linewidth]{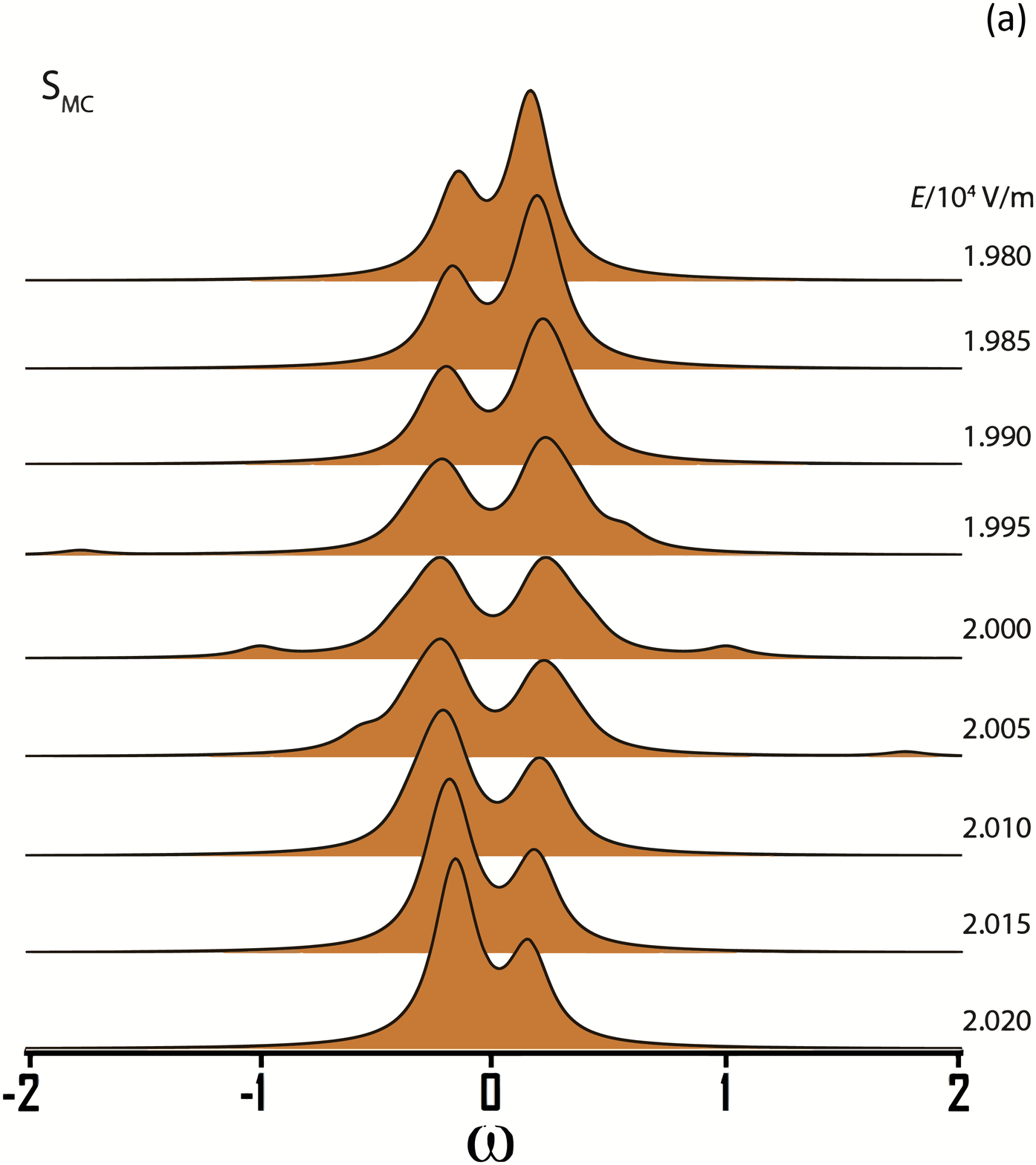}
\includegraphics*[width=0.3\linewidth]{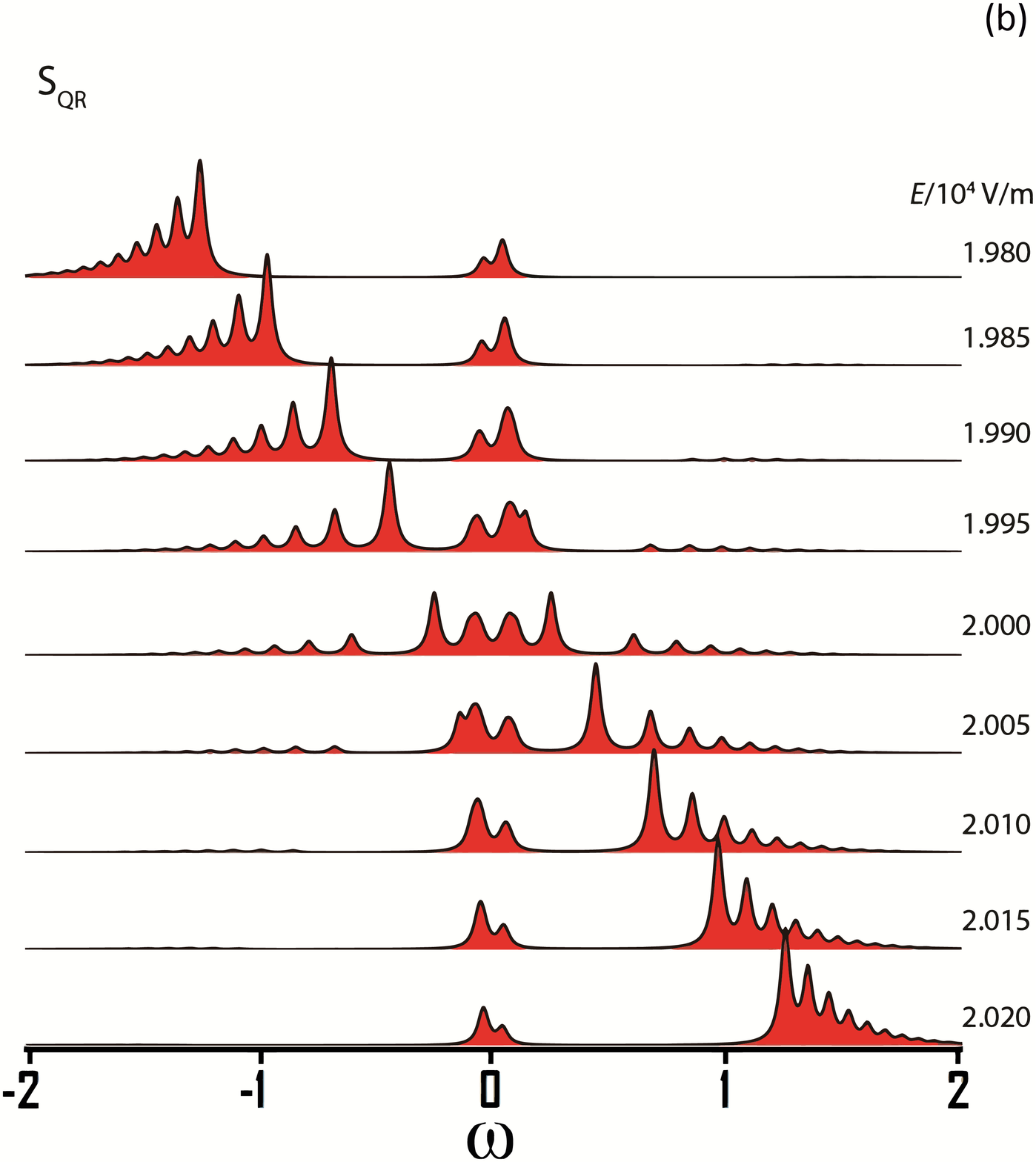}
\caption{(Colour online) Anticrossing in the emission spectrum of the QR-microcavity system at various magnitudes of the external lateral electric field $E$ from $1.98 \times 10^{4} \mathrm{V/m}$ to $2.02 \times 10^{4} \mathrm{V/m}$ with the increment $50 \mathrm{V/m}$: (a) microcavity emission spectrum (brown), (b) direct QR emission spectrum (red). The magnetic flux piercing the QR $\Phi=\Phi_{0}/2$. The resonance case $\Delta=\hbar \omega_{MC}$ corresponds to $E=2.00 \times 10^{4} \mathrm{V/m}$. The microcavity pumping rate $P_{MC}/{\cal G}=0.095$. The emission frequencies are normalised by the QR-microcavity coupling constant ${\cal G}/ \hbar$ and centred around $\omega_{MC}$.}
\label{ES_E_change_F_05}
\end{figure}
A different type of emission spectrum can be observed away from the resonance. This can be achieved for the same system by changing the magnitude of the lateral electric field. In Figs.~\ref{ES_E_change_F_00}--\ref{ES_E_change_F_05} we plot $S_{MC}$ and $S_{QR}$ when $\Delta \ne \hbar \omega_{MC}$ for several values of $E$. Fig.~\ref{ES_E_change_F_00} corresponds to $\Phi=0$, whereas Fig.~\ref{ES_E_change_F_05} corresponds to  $\Phi=\Phi_{0}/2$. Due to the fact that there are non-zero probabilities of finding the system in states with different $N$, the emission spectrum has a pronounced multiplet structure. The microcavity pumping rate is taken as $P_{MC}/ {\cal G}=0.095$. One can clearly see the avoided crossings in the plotted emission spectra, manifesting that the system is in the strong coupling regime. When $\Phi=\Phi_{0}/2$ and the detuning between $\Delta$ and $\hbar \omega_{MC}$ is of the order of ${\cal G}$, the direct QR emission spectrum has the most intensive peaks at the frequencies close to $\omega=\Delta/ \hbar$. This indicates that the QR is almost uncoupled from the microcavity. The more pronounced changes in the emission spectra in Fig.~\ref{ES_E_change_F_05} compared to Fig.~\ref{ES_E_change_F_00} can be explained by different dependences of the energy gap $\Delta$ on the magnitude of the lateral electric field $E$: when $\Phi=\Phi_{0}/2$ the dependence is liner in $E$ and when $\Phi=0$ the dependence is quadratic in $E$.

\begin{figure}[h]
\centering
\includegraphics*[width=0.3\linewidth]{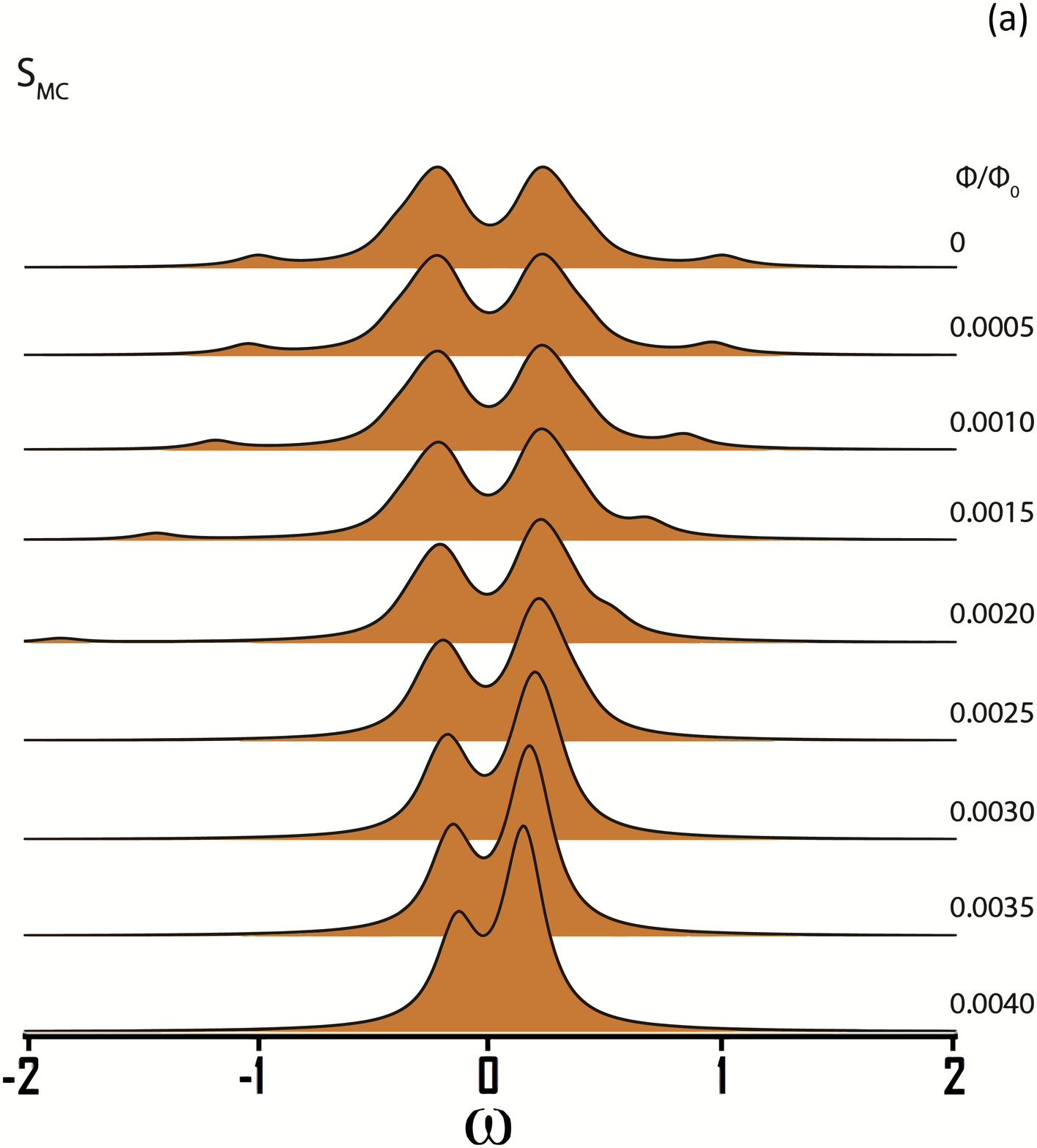}
\includegraphics*[width=0.3\linewidth]{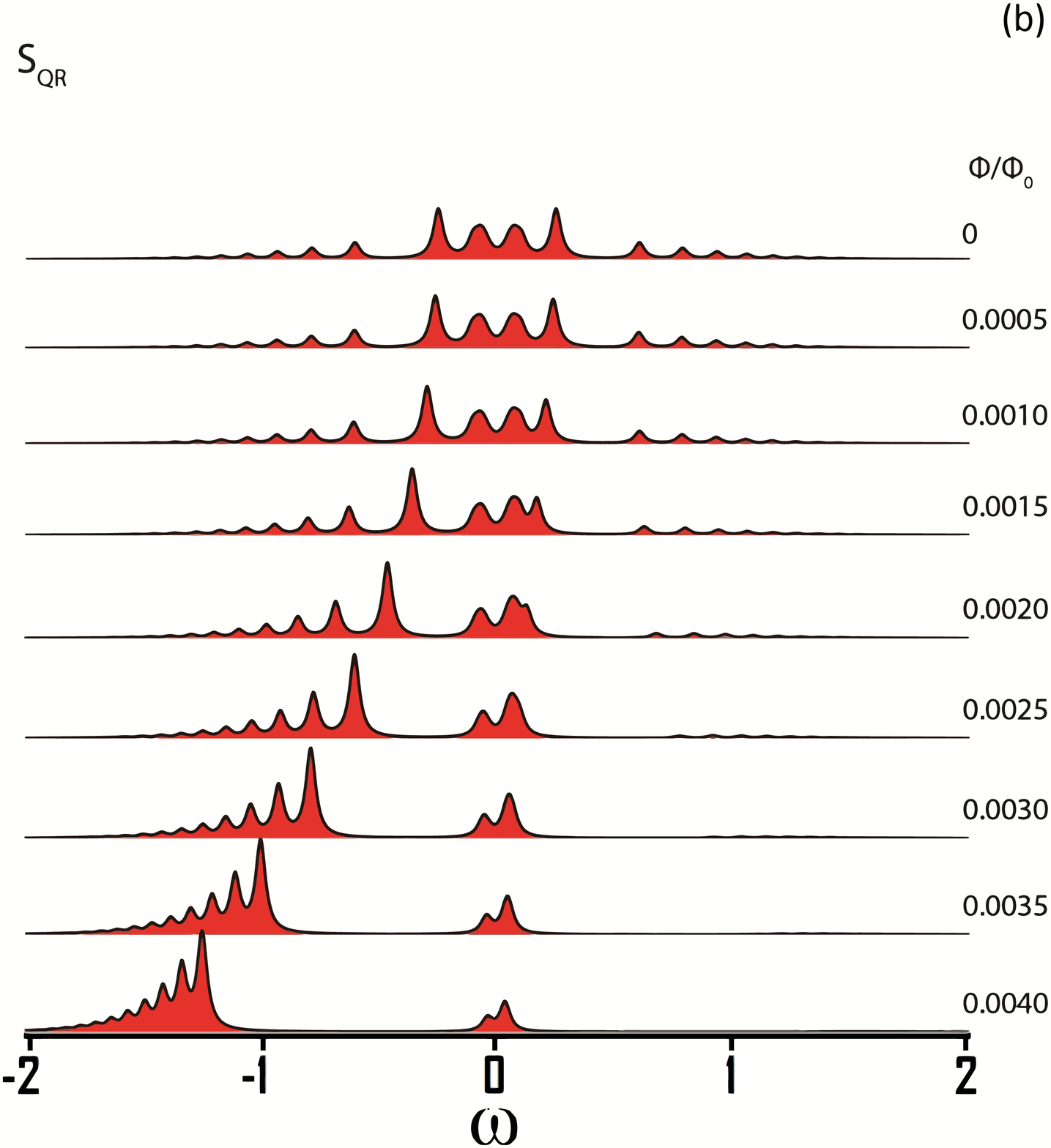}
\caption{(Colour online) Anticrossing in the emission spectrum of the QR-microcavity system at various magnitudes of the magnetic flux $\Phi$ piercing the QR from $0$ to $0.004 \Phi_{0}$ with the increment $5 \times 10^{-4} \Phi_{0}$ and in the presence of the lateral electric field $E=2.00 \times 10^{4} \mathrm{V/m}$: (a) microcavity emission spectrum (brown), (b) direct QR emission spectrum (red). The resonance case $\Delta=\hbar \omega_{MC}$ corresponds to $\Phi=0$. The emission frequencies are normalised by the value of the QR-microcavity coupling constant calculated for $\Phi=0$ (${\cal G}_{0}$) and centred around $\omega_{MC}$. The microcavity pumping rate $P_{MC}/{\cal G}_ {0}=0.095$.}
\label{ES_F_change}
\end{figure}
For a nearly zero flux through the QR, a small change of the flux results in significant changes in $S_{MC}$ and $S_{QR}$, as the presence of a weak magnetic field affects strongly both the QR gap $\Delta$ and the QR-microcavity coupling constant ${\cal G}$. The dependence of the QR gap $\Delta$ on the magnetic flux $\Phi$ piercing the QR can be seen from Fig.~\ref{QREnergySpectrum}, while the QR-microcavity coupling constant ${\cal G}$ magnetic flux dependence can be easily calculated using Eqs.~(\ref{tmatrixelement_ef})--(\ref{d_p}) and Eq.~(\ref{cc_g}). In Fig.~\ref{ES_F_change} we plot $S_{MC}$ and $S_{QR}$ for several values of $\Phi$ near zero. The microcavity pumping rate is taken as $P_{MC}/{\cal G}_{0}=0.095$, where ${\cal G}_{0}$ denotes the value of the QR-microcavity coupling constant for $\Phi=0$. The plotted emission spectra incorporate both the anticrossing behaviour due to detuning of the QR transition energy from the energy of the microcavity mode and the changes in the multiplet structure owing to varying the QR-microcavity coupling strength.

\begin{figure}[h]
\centering
\includegraphics*[width=0.6\linewidth]{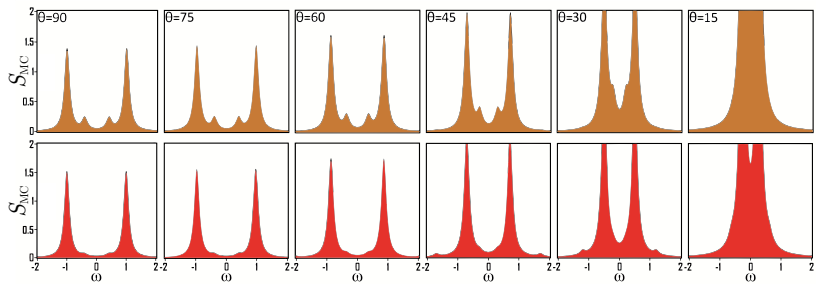}
\caption{(Colour online) Emission spectrum of the QR-microcavity system when the lateral electric field $E=2.00 \times 10^{4} \mathrm{V/m}$ is rotated. The angle $\theta$ is counted between $\mathbf{E}$ and the projection of the microcavity mode polarization vector onto the QR plane $\mathbf{p}$. The upper row (brown) corresponds to the microcavity emission and the lower row (red) correspond to the direct QR emission. The system is in resonance, $\Delta= \hbar \omega_{MC}$. The emission frequencies are normalised by the value of the QR-microcavity coupling constant for $\theta=\pi/2$ (${\cal G}_{\pi/2}$) and centred around $\omega_{MC}$. The microcavity pumping rate $P_{MC}/{\cal G}_{\pi/2}=0.095$.}
\label{ES_angle_change}
\end{figure}
Finally, we calculate the emission spectrum of the QR-microcavity system altering the angle $\theta$ between the direction of the applied electric field and the projection of the microcavity mode polarization vector onto the QR plane. Again, the magnetic flux piercing the QR is either $\Phi=0$ or $\Phi=\Phi_{0}/2$. The system is in the resonance, $\Delta=\hbar \omega_{MC}$. The microcavity pumping rate is taken as $P_{MC}/{\cal G}_{\pi/2}=0.005$, where ${\cal G}_{\pi/2}$ denotes the value of the QR-microcavity coupling constant for $\theta=\pi/2$. The results are plotted in Fig.~\ref{ES_angle_change}. One can see that as the angle $\theta$ is changed, the emission peaks shift towards the microcavity eigenfrequency $\omega_{MC}$, which can be explained by reducing the coupling strength ${\cal G}$. This effect provides an additional way to control the frequency of the satellite peaks in the QR-microcavity emission spectrum and allows a purely spectroscopic measurement of the pump polarization.

In this work we dealt exclusively with the QR inter-subband transitions. However, a similar analysis should be possible for inter-band optical transitions, for which matrix elements and energies can also be tuned by the external fields much easier than in the widely studied QD systems.

In conclusion, we have analyzed the emission spectrum of an Aharonov-Bohm quantum ring placed into a single-mode quantum microcavity. We have shown that the emission spectrum in the strong coupling regime has a multiplet structure and can be tuned by the variation of the magnetic field piercing the quantum ring and by changing the strength and direction of the applied lateral electric field. Thus, it is demonstrated that a microcavity with an embedded QR is a promising system for use as a tunable optical modulator in the THz range. The QR-microcavity system, which allows manipulation of quantum states with  external fields, might also prove to be useful for investigating dephasing mechanisms and for engineering and exploring enhanced light-matter interactions for novel quantum investigations.

\begin{acknowledgements}
We thank Ivan Savenko and Oleg Kibis for valuable discussions and Charles Downing for a critical reading of the manuscript. This work was supported by the FP7 Initial Training Network Spin-Optronics (Grant No. FP7-237252) and FP7 IRSES projects SPINMET (Grant No. FP7-246784) and QOCaN (Grant No. FP7-316432). I.A.S. acknowledges support from Rannis ``Center of excellence in polaritonics'' and FP7 IRSES project POLATER.
\end{acknowledgements}


\newpage

\end{document}